\documentclass[conference]{IEEEtran}
\IEEEoverridecommandlockouts
\usepackage{geometry}
\usepackage{cite}
\usepackage{amsmath,amssymb,amsfonts}
\usepackage{algorithmic}
\usepackage{graphicx}
\usepackage{textcomp}
\usepackage{xcolor}
\def\BibTeX{{\rm B\kern-.05em{\sc i\kern-.025em b}\kern-.08em
    T\kern-.1667em\lower.7ex\hbox{E}\kern-.125emX}}

\usepackage{mathtools}

\usepackage{steinmetz}
\usepackage{authblk}
\newcommand{\minus}{\scalebox{0.75}[1.0]{$-$}}

\usepackage{caption}
\usepackage{subcaption}

\usepackage{blindtext}

\begin{document}

\title{Rate-Splitting Multiple Access\\ for Enhanced URLLC and eMBB in 6G\\ 
\thanks{This work has been partially supported by the U.K. Engineering and Physical Sciences Research Council (EPSRC) under grant EP/R511547/1.}
}
\author{
	Onur~Dizdar$^{1}$, Yijie~Mao$^{1}$, Yunnuo~Xu$^{1}$, Peiying Zhu$^{2}$ and~Bruno~Clerckx$^{1}$\\
$^{1}$Department of Electrical and Electronic Engineering, Imperial College London\\
$^{2}$Huawei Technologies Canada Co. Ltd., Canada\\
Email: \{o.dizdar, y.mao16, yunnuo.xu19, b.clerckx\}@imperial.ac.uk, peiying.zhu@huawei.com}
\IEEEspecialpapernotice{(Invited Paper)}
\maketitle

\begin{abstract}
Rate-Splitting Multiple Access (RSMA) is a flexible and robust multiple access scheme for downlink multi-antenna wireless networks. RSMA relies on Rate-Splitting (RS) at the transmitter and Successive Interference Cancellation (SIC) at the receivers. In this work, we study the performance of RSMA in the scenarios related with the important core services of New Radio (NR) and 6G, namely, enhanced Ultra-Reliable and Low-Latency (URLLC) and enhanced Mobile Broadband Communications (eMBB). We present the optimal system designs employing RSMA that target short-packet and low-latency communications as well as robust communications with high-throughput under the practical and important setup of imperfect Channel State Information at Transmitter (CSIT) originating from user mobility and latency/delay (between CSI acquisition and data transmission) in the network.
We demonstrate via numerical results that RSMA achieves significantly higher performance than Space Division Multiple Access (SDMA) and Non-Orthogonal Multiple Access (NOMA), and is capable of addressing the requirements for enhanced URLLC and eMBB in 6G efficiently.
\end{abstract}


\section{Introduction}
As discussions on 6G progress, it becomes more evident that the requirements for core services and applications are much stricter than those in New Radio (NR). On the other hand, key enabling technologies of NR, such as Multi-User Multiple-Input Multiple-Output (MU-MIMO) in the form of Space Division Multiple Access (SDMA), encounter problems in satisfying the requirements of NR in real-life applications \cite{wang_2017, sakai_2020, dizdar_2021}. Though attractive in MIMO Gaussian Broadcast Channel (BC) with perfect Channel State Information at the transmitter (CSIT) for achieving the entire Degree-of-Freedom (DoF) region \cite{clerckx_2016}, multi-user linear precoding with treat interference as noise receivers (as used in SDMA/MU-MIMO in 4G/5G) is highly sensitive to CSIT inaccuracy, which leads to severe data rate degradation in practical scenarios. Such problems call for urgent solutions to satisfy the demand for exponentially increasing data rate and accessibility in the next generation communications standards. 

Apart from high data rate applications, 6G targets use cases such as health, industrial automation and intelligent transportation with stricter requirements than those in NR. 
For instance, in intelligent transportation and fully automated driving, cooperation among the cars for collision avoidance, environmental awareness and high density platooning requires strict constraints on reliability and latency. 
In industrial automation, use cases such as factory, process, and power system
automation require very low latency and high reliability \cite{popovski_2018, durisi_2016, chen_2018}. Such requirements imply the need for robust and reliable multiple access methods that can achieve higher spectral efficiency than the existing ones.

Rate-Splitting Multiple Access (RSMA) is a multiple access scheme based on Rate-Splitting (RS) and linear precoding for multi-antenna multi-user communications. RSMA splits user messages into common and private parts, and encodes the common parts into one or several common streams while encoding the private parts into separate private streams. The streams are precoded using the available (perfect or imperfect) CSIT, superposed and transmitted via MIMO or Multi-Input Single-Output (MISO) BC \cite{clerckx_2016}. All receivers first decode the common stream(s), perform Successive Interference Cancellation (SIC) and then decode their respective private streams. Each receiver reconstructs its original message from the part of its message embedded in the common stream(s) and its intended private stream.
The key benefit of RSMA is its capability to manage the interference flexibly by allowing the interference to be partially decoded and partially treated as noise. RSMA has been demonstrated to embrace and outperform existing multiple access schemes, i.e., SDMA, Non-Orthogonal Multiple Access (NOMA), Orthogonal Multiple Access (OMA) and multicasting. Thus, RSMA is a promising enabling technology for 6G \cite{mao_2018, joudeh_2016_2,clerckx_2019,dizdar_2020_2,clerckx_2021}.
\begin{figure}[t!]
	\centerline{\includegraphics[width=3.3in,height=3.3in,keepaspectratio]{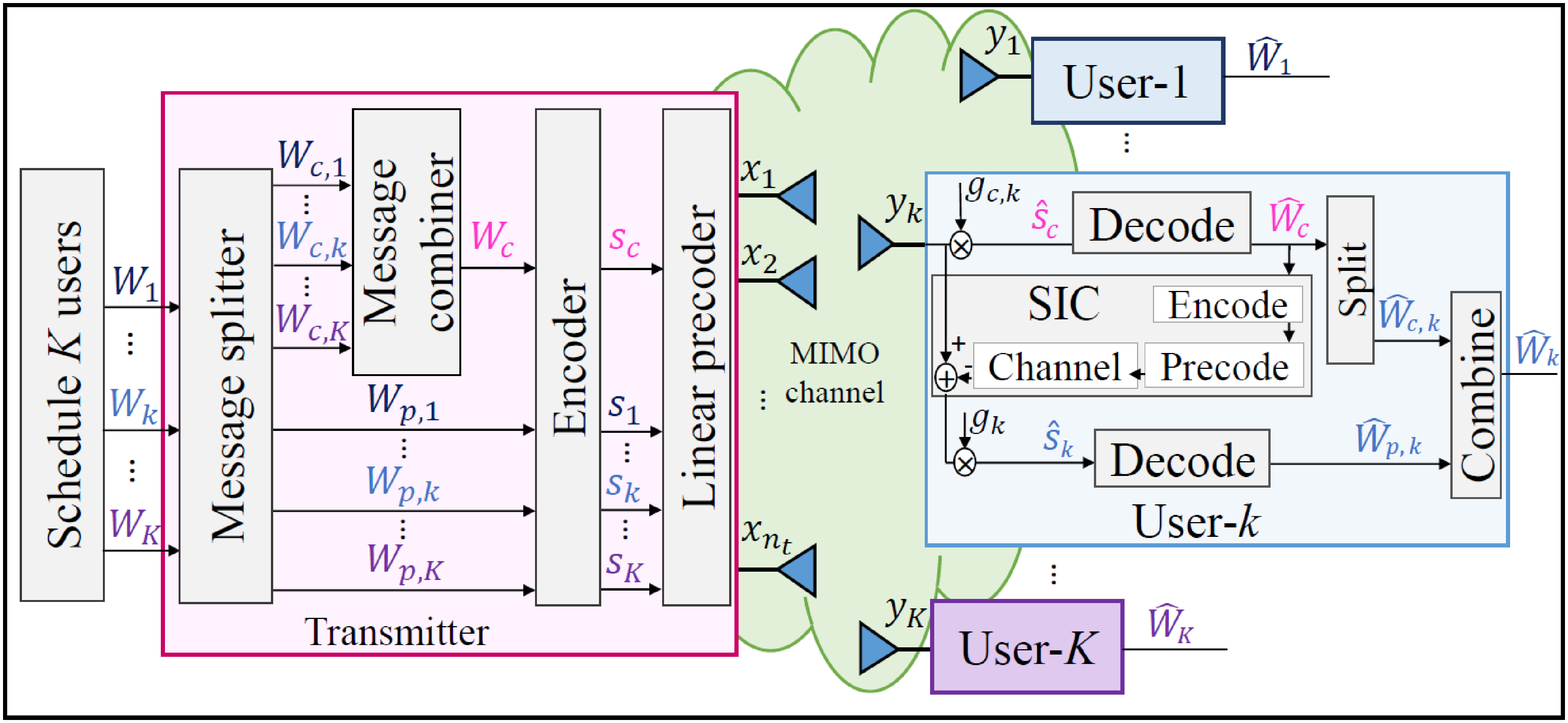}}
	\vspace{-0.2cm}
	\caption{Transmission model of $K$-user 1-layer RSMA}
	\label{fig:system}
	\vspace{-0.7cm}
\end{figure}

In this paper, we present recent results that address the problems and possible solutions in Ultra-Reliable and Low-Latency (URLLC) and enhanced Mobile Broadband Communications (eMBB) core services using RSMA. Specifically, we focus on the performance of RSMA in short-packet communications and scenarios with outdated CSIT due to user mobility and delay in the network, which are major problems in the respective core services. We give optimal system designs with RSMA and compare their performance with those of SDMA and NOMA. We show that RSMA achieves reliable performance under considered target scenarios, and thus, is an excellent enabling technology for 6G.

\textit{Notations:} Vectors are denoted by bold lowercase letters. 
Operations $|.|$ and $||.||$ denote the absolute value of a scalar and $l_{2}$-norm of a vector, respectively. 
$\mathbf{a}^{T}$ and $\mathbf{a}^{H}$ denote the transpose and Hermitian transpose of a vector $\mathbf{a}$, respectively. 
$\mathcal{CN}(0,\sigma^{2})$ denotes the Circularly Symmetric Complex Gaussian (CSCG) distribution with zero mean and variance $\sigma^{2}$. The function $\Gamma(x)=\int_{0}^{\infty}t^{x-1}e^{-t}dt$ is the gamma function and $\Gamma^{\prime}(x)$ is the derivative of $\Gamma(x)$ with respect to $x$.
The function $E_{v}(x)=\int_{1}^{\infty}\frac{e^{-tx}}{t^{v}}dt$, $x > 0$, $\ v \in \mathbb{R}$ is the generalized exponential-integral function. The function $Q^{-1}(x)$ is the inverse Q-function with $Q(x)=\frac{1}{2\pi}\int_{x}^{\infty}e^{-t^{2}/2}dt$. 


\section{Rate-Splitting Multiple Access}
\label{sec:system}
RSMA has been studied in several forms, such as splitting all or some of the users' messages, or using one or more layers of common messages for $K$ users. In this work, we only focus on 1-layer RSMA due to space limitation. The framework and numerical results can be further extended to other forms of RSMA specified in \cite{mao_2018}. Fig.~\ref{fig:system} illustrates the transmission model of 1-layer RSMA.
We consider the MISO BC consisting of one transmitter with $n_{t}$ antennas and $K$ single-antenna users indexed by $\mathcal{K}=\{1,\ldots,K\}$.  The message intended for user-$k$, $W_{k}$, is split into common and private parts, i.e., $W_{c,k}$ and $W_{p,k}$, for $k \in \mathcal{K}$. The split messages are assumed to be independent. The common parts of the messages, $W_{c,k}$, are combined into the common message $W_{c}$. The common message $W_{c}$ and the $K$ private messages $W_{p,k}$ are independently encoded into streams $s_{c}$ and $s_{k}$, $k\in\mathcal{K}$, respectively. The streams satisfy $\mathbb{E}\left\lbrace \mathbf{s}\mathbf{s}^{H}\right\rbrace =\mathbf{I}$, where  $\mathbf{s}=\left[s_{c}, s_{1}, \ldots, s_{K}\right]$. The transmit signal is given as
\begin{align}
\mathbf{x}=\mathbf{p}_{c}s_{c}+\sum_{k \in \mathcal{K}}\mathbf{p}_{k}s_{k},
\label{eqn:txsignal}
\end{align}
where $\mathbf{p}_{c}, \mathbf{p}_{k} \in\mathbb{C}^{n_t}$ are the linear precoders applied to the common stream and the private stream of user-$k$, respectively. 

The signal received by user-$k$ is written as
\begin{align}
	y_{k}=\mathbf{h}_{k}^{H}[m]\mathbf{x}+n_{k}, \quad \forall k\in\mathcal{K},  
	\label{eqn:rxsignal}
\end{align}
where $\mathbf{h}_{k}[m] \in \mathbb{C}^{n_{t}}$ is the channel vector between the transmitter and user-$k$ at time instant $m$ and $n_{k} \sim \mathcal{CN}(0,\sigma_{k}^{2})$ is the Additive White Gaussian Noise (AWGN) component. 
The receivers apply SIC to detect the common and their corresponding private streams. The common stream is detected first to obtain the common message estimate $\hat{W}_{c}$ by treating the private streams as noise. 
The Signal-to-Interference-plus-Noise Ratio (SINR) for the common stream at user-$k$ is
\vspace{-0.15cm}
\begin{align}
	\gamma_{c,k}=\frac{|\mathbf{h}_{k}[m]^{H}\mathbf{p}_{c}|^{2}}{\sum_{i \in \mathcal{K}}|\mathbf{h}_{k}[m]^{H}\mathbf{p}_{i}|^{2}+\sigma_{k}^{2}}.
	\label{eqn:common_SINR}
\end{align}

The maximum rate for the common stream is $  R_{c}\hspace{-0.1cm}=\min\left\lbrace\log_{2}(1+\gamma_{c,1}), \ldots, \log_{2}(1+\gamma_{c,K}) \right\rbrace$.
The common stream consists of the common messages intended for all users, i.e, $R_{c}=\sum_{i \in \mathcal{K}}C_{k}$,
where $C_{k}$ is the portion of the rate of the common stream intended for user-$k$.  

The common stream is reconstructed using $\hat{W}_{c}$ and subtracted from the received signal. The remaining signal is used to detect the private message $\hat{W}_{p,k}$. 
The SINR and maximum rate for the private stream at user-$k$ are given by
\begin{align}
	\gamma_{k}=\frac{|\mathbf{h}_{k}[m]^{H}\mathbf{p}_{k}|^{2}}{\sum_{i \in \mathcal{K}, i \neq k}|\mathbf{h}_{k}[m]^{H}\mathbf{p}_{i}|^{2}+\sigma_{k}^{2}},
	\label{eqn:private_SINR}
\end{align}
and $R_{k}=\log_{2}(1+\gamma_{k})$, respectively. 
The overall maximum rate for user-$k$ is $R_{k,\mathrm{total}}=C_{k}+R_{k}$.

\section{RSMA for Enhanced URLLC and eMBB}
The targeted use cases in 6G imply the Key Performance Indicators (KPIs) like high data rate and ultra-reliable and low-latency communications are as important in 6G as they are in NR.
The throughput requirement is expected to be above $1$~Tbps for high data rate applications, and the latency requirement is predicted to be below $0.1$~ms with a transmission reliability down to a Block Error Rate (BLER) of $10^{-9}$. The challenge grows when such requirements are to be supported with mobile users and user speeds up to $1000$~km/h \cite{2,4,5,10, dizdar_2020_2}.  
In this section, we address how the enhanced data rate, latency and reliability requirements of 6G can be satisfied by RSMA in a reliable and robust manner.

\subsection{RSMA for Enhanced URLLC}
\label{sec:urllc}

The most critical sources of latency in cellular networks are identified as retransmissions caused by channel estimation errors and congestion, link establishment, and the minimum data block length. 
A method to reduce the latency due to retransmissions is to employ a robust scheme that operates reliably in the presence of interference. 
Reliable communications is vital especially in multiple access scenarios with low-latency, such as intelligent transportation, where packet losses may result in serious consequences \cite{durisi_2016, chen_2018}. As discussed in Section~\ref{sec:embb}, RSMA was shown to achieve robust performance with improved data rate under interference, and specifically, interference due to user mobility. 

Reducing the transmitted packet size is another method to achieve low-latency communications.
RSMA was first shown to achieve improved performance over SDMA and NOMA with finite length polar codes by Link-Level Simulations (LLS) in \cite{dizdar_2020}. However, the precoders used in \cite{dizdar_2020} are optimized following the approach in \cite{joudeh_2016_2}, which is based on the Shannon capacity formula assuming infinite blocklength channel coding. Thus, the considered precoders are expected to be sub-optimal for Finite Blocklength (FBL) coding. 

The authors investigate the optimal performance of RSMA with short-packets in \cite{xu_2021} using the achievable rate formula for FBL coding given in \cite{polyanskiy_2010}. Accordingly, the rates for the common and private streams with FBL coding are written as
\begin{align}
	R_{c,k}&\approx \log_{2}(1+\gamma_{c,k})-\sqrt{\frac{V_{c,k}}{N_{c}}}Q^{-1}(\xi), \nonumber \\
	R_{p,k}&\approx \log_{2}(1+\gamma_{p,k})-\sqrt{\frac{V_{p,k}}{N_{k}}}Q^{-1}(\xi),
\end{align}
where $R_{c,k}$ and $R_{p,k}$ represent the achievable rates under the target BLER $\xi$, $N_{c}$ and $N_{k}$ represent the blocklength for the common stream and private stream of user-$k$, $\forall k \in \mathcal{K}$, respectively. 
The channel dispersion parameters are expressed as \mbox{$V_{c,k}=(\log_{2}e)^{2}[1-(1+\gamma_{c,k})^{-2}]$} and \mbox{$V_{p,k}=(\log_{2}e)^{2}[1-(1+\gamma_{p,k})^{-2}]$}, $\forall k \in \mathcal{K}$. 
The optimization problem to obtain the precoders which maximize the achievable Weighted Sum Rate (WSR) is written as 
\vspace{-0.1cm}
\begin{subequations}
	\begin{alignat}{3}
		\min_{\mathbf{P}, \mathbf{c}}&     \quad  \sum_{k \in \mathcal{K}}u_{k}\left(C_{k}+ R_{p,k}\right)      \label{eqn:obj_f}   \\
		\text{s.t.}&  \quad  \sum_{k^{\prime} \in \mathcal{K}}C_{k^{\prime}} \leq R_{c,k}, \quad \forall k \in \mathcal{K}  \label{eqn:common_rate_1_f} \\
		& \quad  \mathrm{tr}(\mathbf{P}\mathbf{P}^ {H}) \leq P,  \label{eqn:total_power_f} \\
		& \quad C_{k}+ R_{p,k} \geq r_{k}^{th} , \quad \forall k \in \mathcal{K} \label{eqn:qos} \\
		& \quad   \mathbf{c} \geq \mathbf{0}, \label{eqn:x}
	\end{alignat}
	\label{eqn:problem1_final}
\end{subequations}
\hspace{-0.15cm}where $C_{k}$ represents the portion of the common rate for user-$k$,  $\mathbf{c}=[C_{1}, C_{2}, \ldots, C_{K}]$ and \mbox{$\mathbf{P}=\left[\mathbf{p}_{c},  \mathbf{p}_{1}, \ldots, \mathbf{p}_{K}\right] $}. Constraint \eqref{eqn:common_rate_1_f} ensures that the common stream is successfully decoded by both users. Constraint \eqref{eqn:total_power_f} is the
transmit power constraint and constraint \eqref{eqn:qos} ensures the Quality-of-Service (QoS) rate, with $r_{k}^{th}$ representing the minimum rate of user-$k$. 

A Successive Convex Approximation (SCA) based algorithm is proposed in \cite{xu_2021} to deal with the non-convex problem \eqref{eqn:problem1_final}. 
We refer the interested reader to \cite{xu_2021} for details on the proposed algorithm. 
It will be shown in Section~\ref{sec:results} that RSMA achieves identical sum-rate performance with FBL coding using shorter codes than SDMA and NOMA, making it a suitable enabling technology for enhanced URLLC.

\subsection{RSMA for eMBB with Enhanced Data Rate}
\label{sec:embb}
It has been discussed in \cite{dizdar_2020_2} that the path to achieving a higher data rate in 6G starts with addressing the problems affecting the performance of data services in NR. It is well-known that the enabling technologies for high data rate, such as (massive) MIMO, is vulnerable to CSIT imperfections which result in interference in multi-user transmissions. 
Several causes of interference have been reported so far to limit the achievable rate in downlink (DL) multi-user transmissions in NR.
An example is the interference due to user blockage in multi-user transmission, where one of the users served in the DL goes into outage and is unable to update its Channel State Information (CSI) report, resulting in rate loss for all users \cite{sakai_2020}. Another example is the separate implementation of the Physical (PHY)-layer design (such as precoder design) and the scheduling algorithm of the multiple access scheme. Such implementation procedure creates a rate loss to those multiple access schemes whose performance heavily depends on the channel disparities and orthogonalities of the users being served, such as SDMA and NOMA. 

The robustness of RSMA to interference and its superior performance compared to SDMA, NOMA and OMA in multi-antenna settings have been shown in numerous studies 
\cite{dai_2016, joudeh_2016, joudeh_2016_2, hao_2015, mao_2018, clerckx_2019, mao_2019_2}.
The effects of user grouping and scheduling have also been studied in\cite{dai_2016, mao_2018, mao_2019_2}, where RSMA has been shown to be more robust to the interference resulting from suboptimal grouping compared to SDMA and NOMA.

Apart from the abovementioned causes of interference, the (arguably) most important cause of interference in NR originates from user mobility. Combined with the delay in CSI acquisition/feedback from mobile users, mobility results in outdated CSIT due to the variations in the wireless medium.  
One of the earliest problems reported in field trials of NR is the limited performance of MU-MIMO communications as a result of outdated CSIT due to mobility. The processing delay in the system is reported to become larger than the coherence time of the channel even when the users are moving with a speed of $30$km/h, making the CSIT useless for multi-user beamforming \cite{wang_2017}.
It is expected that the effects of mobility will be encountered more frequently and severely in 6G, and thus, the problem calls for an urgent solution. 

The performance of RSMA was studied in \cite{dizdar_2021} under imperfect CSIT due to user mobility and latency in multi-user (massive) MIMO systems. The authors consider the MISO BC setting specified in Section~\ref{sec:system}. The users in the system are mobile with arbitrary speed. The transmitter employs Zero-Forcing (ZF) precoders for the private streams and a random beamformer for the common stream. The transmit signal for the considered system model is written as in \eqref{eqn:txsignal} with $\mathbf{p}_{c}=\sqrt{P(1-t)} \mathbf{f}_{c}$ and $\mathbf{p}_{k}=\sqrt{\frac{Pt}{K}}\mathbf{f}_{k}$,
where $\mathbf{f}_{c}, \mathbf{f}_{k}\in\mathbb{C}^{n_t}$ satisfy $||\mathbf{f}_{c}||^{2}=1$ and $||\mathbf{f}_{k}||^{2}=1$, $\forall k \in \mathcal{K}$ and $P$ is the total transmit power. 
The power allocation coefficient $t$ determines the distribution of power among the common and the private precoders, and effectively controls the splitting of user rates.  

The signal received by user-$k$ is written as in \eqref{eqn:rxsignal}, with the channel coefficients at time instant $m$ expressed as 
\begin{align}
	\mathbf{h}_{k}[m]=\sqrt{\epsilon^{2}} \mathbf{h}_{k}[m\minus1] + \sqrt{1-\epsilon^{2}} \mathbf{e}_{k}[m],
\end{align}
where $\mathbf{h}_{k}[m]$ denotes the spatially uncorrelated Rayleigh flat fading channel with i.i.d. entries according to $\mathcal{CN}(0,1)$, $\epsilon=J_{0}(2\pi f_{D}T)$ denotes the time correlation coefficient obeying the Jakes' model and $\mathbf{e}_{k}[m]$ has i.i.d. entries according to $\mathcal{CN}(0,1)$ \cite{kim_2011}. The parameter $f_{D}=vf_{c}/c$ is the maximum Doppler frequency for given carrier frequency $f_{c}$ and user speed $v$ ($c$ is the speed of light) and $T$ is the channel instantiation interval ($T$ models the CSI report delay in seconds). 
In words, $\mathbf{h}_{k}[m]$ is the instantaneous channel vector observed at time instant $m$. However, the transmitter only has knowledge of the channel vector observed at time instant $m\minus1$, {\sl i.e.,} $\mathbf{h}_{k}[m\minus 1]$, due to the latency in CSI report from each user-$k$. The transmitter uses the CSIT $\mathbf{h}_{k}[m\minus 1]$ at time instant $m$ to calculate the precoders. The random beamformer for the common precoder is assumed to be independent of $\mathbf{h}_{k}[m\minus 1]$, $\mathbf{e}_{k}[m]$, and $\mathbf{f}_{k}$, $\forall k\in\mathcal{K}$.
The ergodic rates of the common and private streams are expressed as 
\vspace{-0.15cm}
\begin{align}
	R_{c}(t)&\hspace{-0.1cm}=\hspace{-0.1cm}\mathbb{E}\left\lbrace \hspace{-0.1cm}\log_{2}\hspace{-0.1cm}\left(\hspace{-0.1cm} 1\hspace{-0.1cm}+\hspace{-0.1cm}\min_{k\in\mathcal{K}}\left[\frac{P(1-t)|\mathbf{h}^{H}_{k}[m]\mathbf{f}_{c}|^{2}}{\hspace{-0.1cm}1\hspace{-0.1cm}+\hspace{-0.1cm}\frac{Pt}{K}\sum_{j\in\mathcal{K}}|\mathbf{h}^{H}_{k}[m]\mathbf{f}_{j}|^{2}} \right] \right) \hspace{-0.1cm}\right\rbrace, \nonumber \\ 
	R_{k}(t)&\hspace{-0.1cm}=\hspace{-0.1cm}\mathbb{E}\left\lbrace \hspace{-0.1cm}\log_{2}\hspace{-0.1cm}\left(\hspace{-0.1cm} 1\hspace{-0.1cm}+\hspace{-0.1cm} \frac{\frac{Pt}{K} |\mathbf{h}^{H}_{k}[m]\mathbf{f}_{k}|^{2}}{1+ \frac{Pt}{K} \sum_{j\in\mathcal{K}, j\neq k}|\mathbf{h}^{H}_{k}[m]\mathbf{f}_{j}|^{2}} \right) \hspace{-0.1cm}\right\rbrace,
	\label{eqn:rates}
\end{align}
where the expectations are defined over the user channels $\mathbf{h}_{k}$ and $\sigma_{k}^{2}=1$, $\forall k \in \mathcal{K}$, without loss of generality.
The ergodic sum-rate is written as
	\vspace{-0.2cm}
\begin{align}
	R_{\mathrm{RSMA}}(t)=R_{c}(t)+\sum_{k=1}^{K}R_{k}(t).
	\label{eqn:sumrate}
\end{align}
	\vspace{-0.3cm}
	
As seen from expression \eqref{eqn:sumrate}, the ergodic sum-rate is dependent on the power allocation coefficient $t$. A natural question arises on how to set the coefficient $t$ optimally to achieve the maximum sum-rate with low-complexity, so that it can be calculated easily in practical systems. For this purpose, the authors in \cite{dizdar_2021} derive a tractable lower bound for the ergodic sum-rate and obtain a closed-form solution for the optimal power allocation coefficient $t$ that maximizes the obtained lower bound. 
It is shown that an approximation $\widetilde{R}_{\mathrm{RSMA}}(t)$ of the expression \eqref{eqn:sumrate} can be lower-bounded as  
\vspace{-0.2cm}
\begin{align}
	&\widetilde{R}_{\mathrm{RSMA}}(t) \geq \log_{2}\left(1+P(1-t)e^{\beta} \right) \nonumber \\
	&+\hspace{-0.1cm}K\hspace{-0.05cm}\log_{2}\hspace{-0.1cm}\left(\hspace{-0.1cm}1\hspace{-0.08cm}+\hspace{-0.08cm}\frac{P}{K}e^{\mu}t\hspace{-0.05cm}\right) \hspace{-0.08cm}-\hspace{-0.08cm}K\hspace{-0.05cm}\log_{2}\hspace{-0.1cm}\left(\hspace{-0.1cm}1\hspace{-0.1cm}+\hspace{-0.1cm}\frac{(K-1)(1-\epsilon^{2})P}{K}t\hspace{-0.05cm}\right)
	\label{eqn:lowerbound}
\end{align}
where $\mu\triangleq\ln(\theta)+\Gamma^{\prime}(D)/\Gamma(D)$, $\beta\triangleq -\gamma-\ln(K)-e^{\frac{K^{2}}{P\theta t}}\sum_{m=1}^{\left \lfloor{DK}\right \rceil}\mathrm{E}_{m}\left(\frac{K^{2}}{P\theta t}\right)$,
$\gamma\approx0.577$ is the Euler-Mascheroni constant, $D\triangleq\frac{\left[ \epsilon^{2}(n_{t}+1)+(1-2\epsilon^{2})K\right] ^{2}}{\epsilon^{4}(n_{t}+1)+(1-2\epsilon^{2})K}$ 
and \mbox{$\theta\triangleq\frac{\epsilon^{4}(n_{t}+1)+(1-2\epsilon^{2})K}{\epsilon^{2}(n_{t}+1)+(1-2\epsilon^{2})K}$.}
Under the assumption of $Pt \rightarrow \infty$, a closed-form solution for the value of $t$ that maximizes the lower bound in \eqref{eqn:lowerbound} is found as 
\begin{align}
	t_{opt}=
	\left\{
	\begin{array}{ll}
		\frac{\rho(K-1)}{\rho(\omega+K)-K},  &  \mbox{if} \ \rho(\omega+1)/K > 1\\
		\hspace{0.8cm} 1\hspace{0.6cm}, & \mbox{otherwise},
	\end{array}
	\right.
	\label{eq:poweralloc}
\end{align} 
with $\omega\triangleq \frac{(K-1)(1-\epsilon^{2})P}{K}$ and $\rho \triangleq\frac{K}{\theta(\left \lfloor{DK}\right \rceil-1)}e^{-\gamma-\frac{1}{2(\left \lfloor{DK}\right \rceil-1)}}$. 

We refer the interested reader to \cite{dizdar_2021} for more details on the derivations of the lower bound \eqref{eqn:lowerbound} and the closed-form expression \eqref{eq:poweralloc} for $t_{opt}$. As mentioned before, the optimal power allocation coefficient $t_{opt}$ in \eqref{eq:poweralloc} controls the splitting of user rates depending on the parameters $n_{t}$, $K$, $P$ and $\epsilon$. It will be shown in Section~\ref{sec:results} that RSMA used with the power allocation coefficient $t_{opt}$ achieves a robust sum-rate and throughput performance under various user speeds and outperforms SDMA significantly, for which the throughput saturates with increasing SNR. The results show that RSMA ensures robust multi-user connectivity under user mobility with high data rates, and thus, is an excellent candidate to satisfy the data rate requirements of 6G.

\section{Numerical Results}
\label{sec:results}
In this section, we investigate the performance of RSMA with FBL codes and under user mobility as described in Sections~\ref{sec:urllc} and \ref{sec:embb}.  
\subsection{Performance with Short-Blocklength Codes}
\begin{figure}[t!]
	\centerline{\includegraphics[width=3.9in,height=3.9in,keepaspectratio]{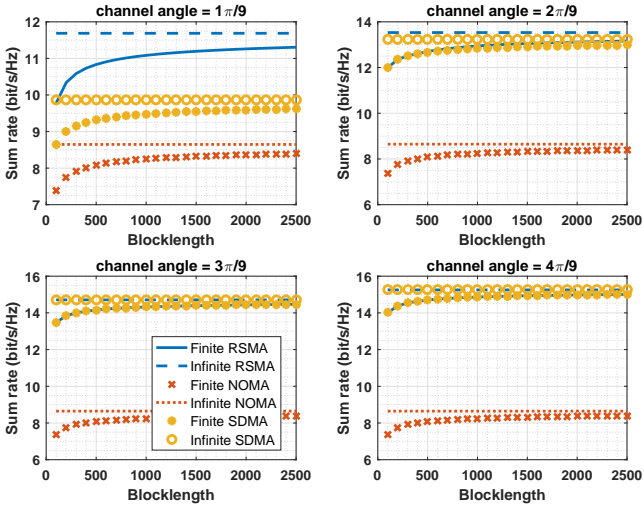}}
	\caption{WSR vs. blocklength}
	\label{fig:FBL}
	\vspace{-0.5cm}
\end{figure}

We perform our analysis over a two-user scenario with $n_{t}=4$. The user channels are realized as \mbox{$\mathbf{h}_{1}=[1,\ 1,\ 1,\ 1]^{H}$} and \mbox{$\mathbf{h}_{2}=[1,\ e^{j\theta},\ e^{j2\theta},\ e^{j3\theta}]^{H}$}, where $\theta \in \left[0, \frac{\pi} {2}\right] $ represents the angle between user channels. 
Fig.~\ref{fig:FBL} gives the sum-rate performance of RSMA, SDMA and NOMA with respect to blocklength for $\theta=\left\lbrace \frac{\pi}{9}, \frac{2\pi}{9}, \frac{3\pi}{9}, \frac{4\pi}{9} \right\rbrace  $. We consider an equal weighting between users, {\sl i.e.,} $\mathbf{u}=[1, 1]^{T}$ and the QoS minimum rate is set as $r_{k}^{th}=0$, $\forall k \in \mathcal{K}$. The SNR is set to $20$dB. The target BLER for the system is set as $10^{-5}$. Since RSMA and NOMA employ SIC for message detection and decoding, the corresponding target BLERs are set as $\xi_{\mathrm{RSMA}}=\xi_{\mathrm{NOMA}}=5\times10^{-6}$ to ensure that the approximated overall BLER does not exceed $10^{-5}$. The target BLER for SDMA is set as $\xi_{\mathrm{SDMA}}=10^{-5}$. 

The notations \textit{infinite} and \textit{finite} in the figure respectively represent the performance with infinite blocklength and with finite blocklength where $N_{c}=N_{k}$, $\forall k \in \mathcal{K}$. 
It can be observed from Fig.~\ref{fig:FBL} that the WSRs of all three strategies increase with blocklength and approach the performance with infinite blocklength, as expected. 
RSMA outperforms SDMA and NOMA in all considered scenarios and the performance gain of RSMA becomes significant especially when $\theta=\frac{\pi}{9}$, for which RSMA only requires a blocklength of $100$ bits to achieve a WSR of $9.7$ bits/s/Hz, while SDMA requires a significantly longer blocklength of around $2500$ bits. 
The performance of RSMA is almost identical to that of
SDMA for other considered values of $\theta$, and both achieve better WSR performance than
NOMA. 

The numerical results show that RSMA can achieve the same WSR with equal or shorter blocklengths than SDMA and NOMA. We can conclude that RSMA has the potential to reduce the latency in a reliable manner for the core services and use cases with low-latency in 6G.

\subsection{Performance under User Mobility}
We study the sum-rate and throughput performance of SDMA and RSMA with the proposed power allocation algorithm by Monte-Carlo Simulations and LLS. We use a transceiver architecture for RSMA with finite alphabet modulation schemes, finite-length channel coding, and an Adaptive Modulation and Coding (AMC) algorithm, as described in \cite{dizdar_2020}. The AMC algorithm selects a suitable modulation-coding rate pair for transmission depending on the transmit rates. The transmit rate calculations for AMC are performed assuming $\mathbf{h}_{k}[m]$, $\forall k \in \mathcal{K}$ are known in the AMC module (only for modulation-coding scheme selection). We employ the modulation schemes $4$-QAM, $16$-QAM, $64$-QAM and $256$-QAM and polar coding. The reader is referred to \cite{dizdar_2020} for details on the architecture and LLS platform.

We investigate the performance of SDMA and RSMA with the proposed power allocation algorithm in \eqref{eq:poweralloc} using Cyclic-Prefix - Orthogononal Frequency Division Multiplexing (CP-OFDM) waveforms in realistic channel models. The OFDM subcarrier spacing is set as $30$kHz. Data is carried over $256$ subcarriers. The CP length is $10\mu$s. We assume the system is operating at $3.5$GHz and the CSI acquisition delay is $10$ms. We calculate and apply precoders separately for each subcarrier. The coefficient $t_{opt}$ is constant for all subcarriers since $n_{t}$, $K$, $P$ and $\epsilon$ do not vary with subcarrier indexes.
We use the Quadriga Channel Generator \cite{jaeckel_2019} to generate channels according to the 3GPP Urban Macro-Cell channel model. The channels of each user have a delay spread of $300$ns and $23$ clusters, with each cluster consisting of $20$ rays.

\begin{figure}[t!]
	\centering
	\begin{subfigure}[t]{.25\textwidth}
		\centerline{\includegraphics[width=2.0in,height=2.0in,keepaspectratio]{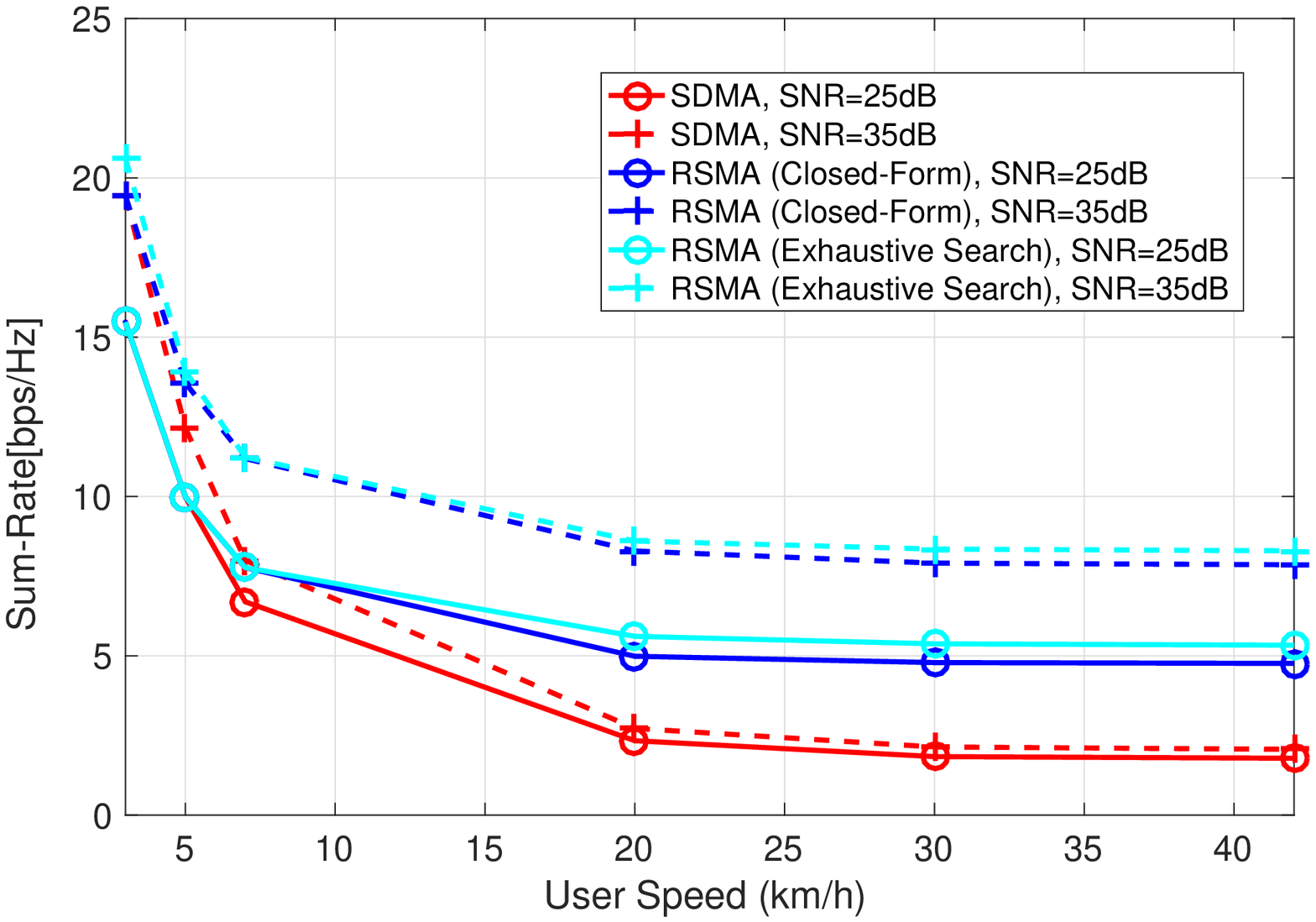}}
		\caption{Sum-rate vs. user speed.}
		\label{fig:cap_doppler_quadriga}
	\end{subfigure}%
	\begin{subfigure}[t]{.25\textwidth}
		\centerline{\includegraphics[width=1.9in,height=1.90in,keepaspectratio]{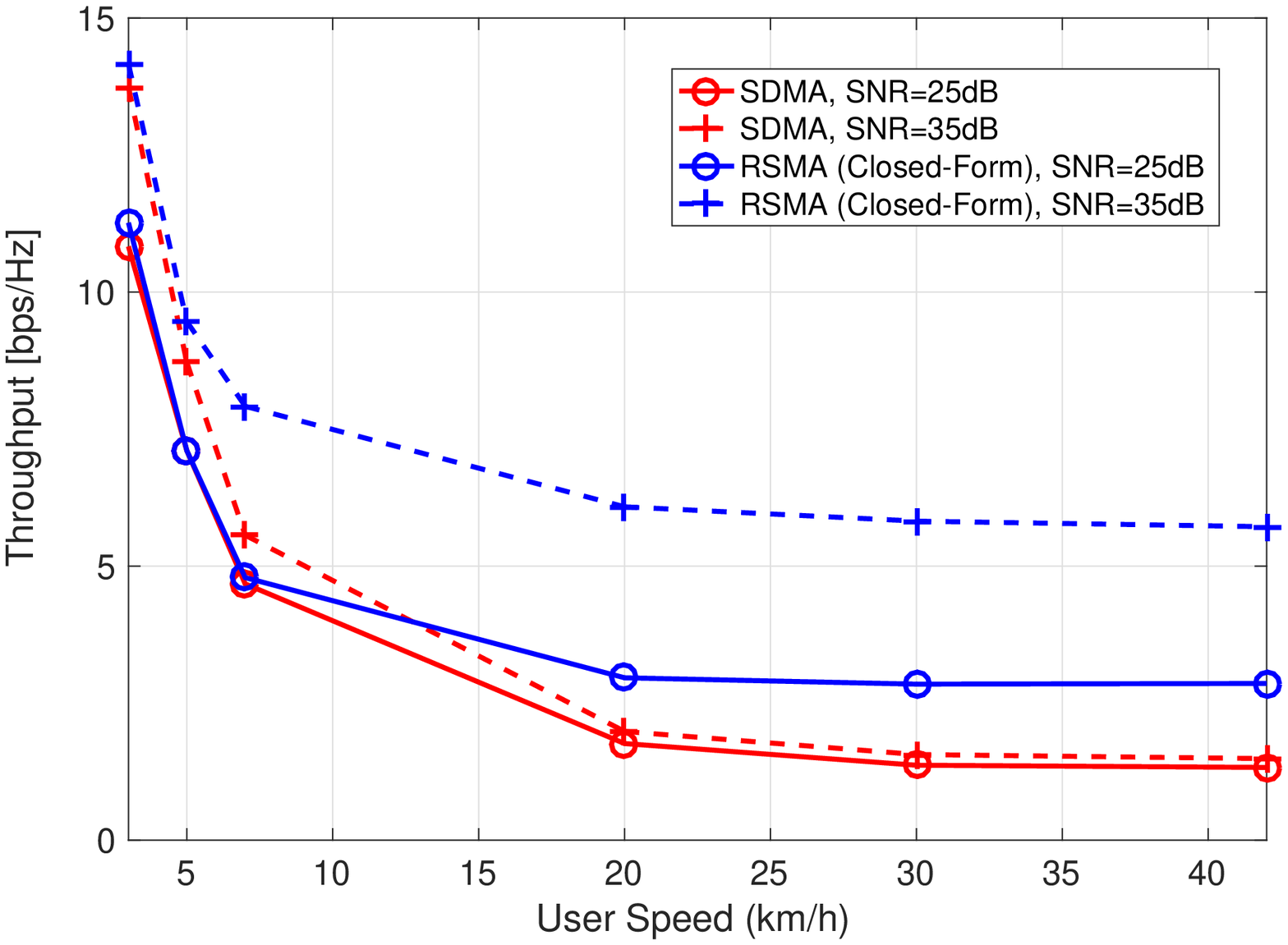}}
		\caption{Throughput vs. user speed.}
		\label{fig:tp_doppler_quadriga}
	\end{subfigure}
	\caption{Sum-rate and throughput vs. user speed with OFDM and 3GPP channel model.}
	\vspace{-0.5cm}
\end{figure}

\begin{figure}[t!]
	\centering
	\begin{subfigure}[t]{.25\textwidth}
		\centerline{\includegraphics[width=1.9in,height=1.3in]{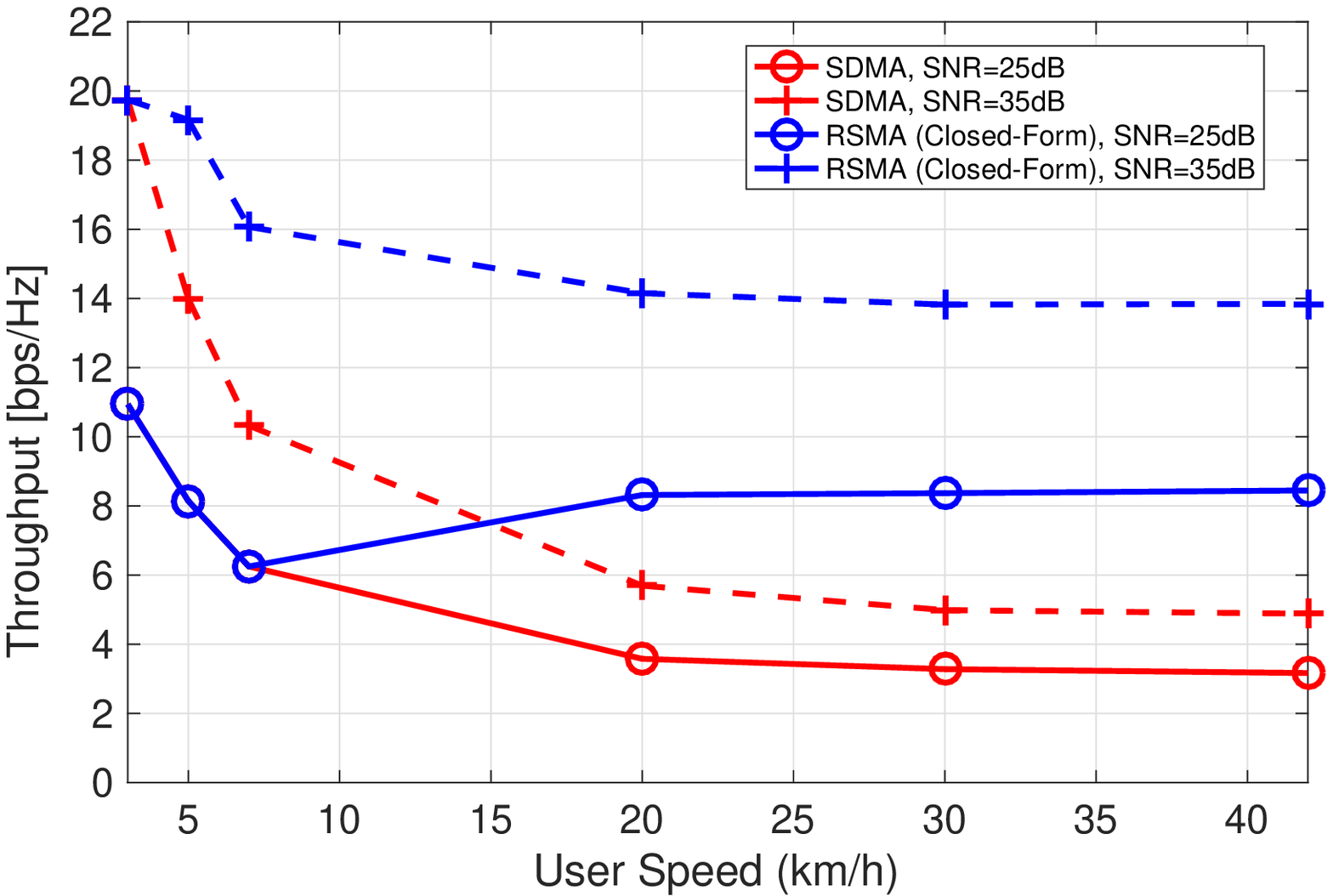}}
		\caption{Throughput vs. user speed,\\ $Q_{c}=Q_{p}=2$.}
		\label{fig:tp_doppler_qc2qp2_quadriga}
	\end{subfigure}%
	\begin{subfigure}[t]{.25\textwidth}
		\centerline{\includegraphics[width=1.9in,height=1.3in]{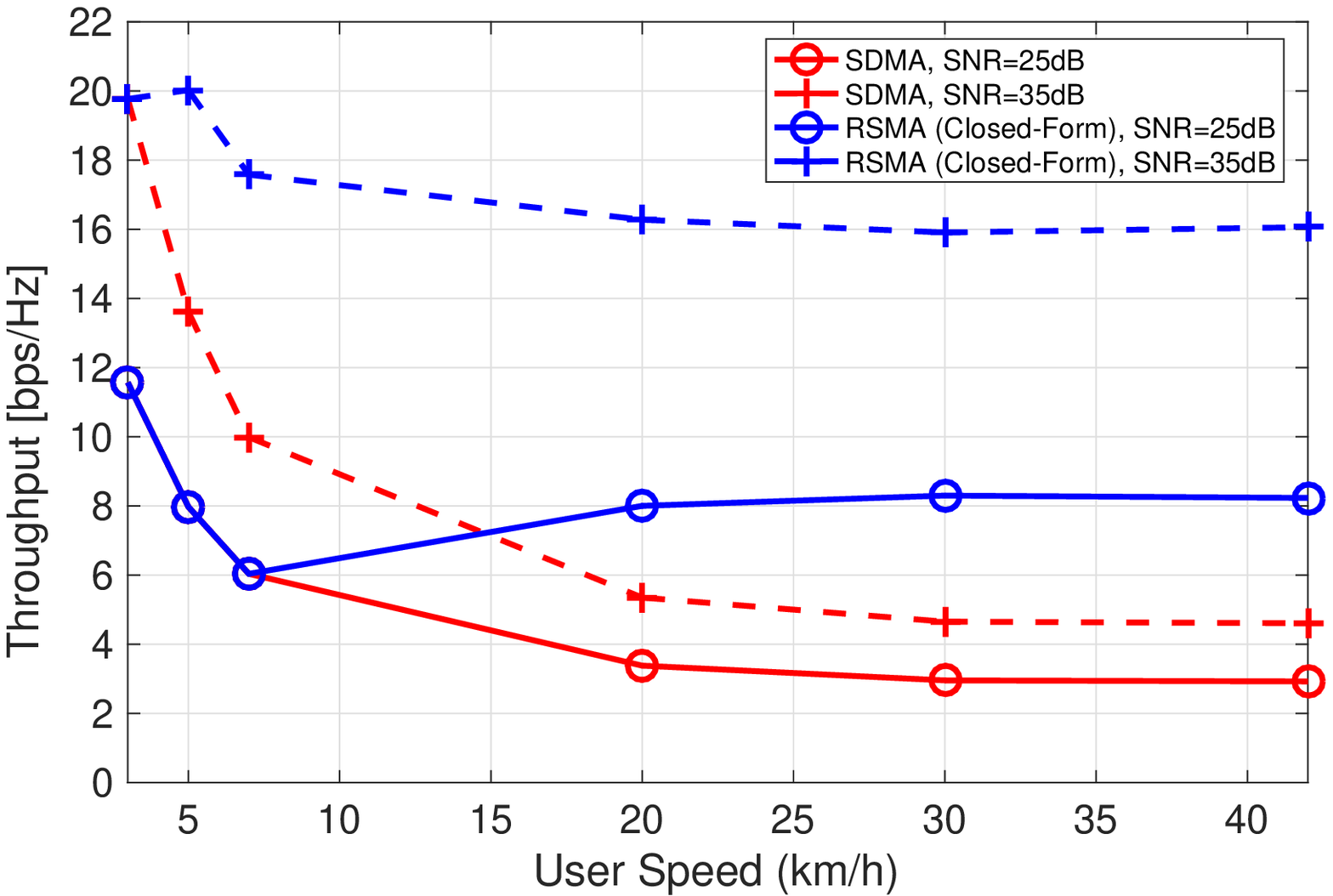}}
		\caption{Throughput vs. user speed, \\$Q_{c}=Q_{p}=3$.}
		\label{fig:tp_doppler_qc3qp3_quadriga}
	\end{subfigure}
	\caption{Throughput vs. user speed with OFDM and 3GPP channel model, multiple receive antennas.}
	\vspace{-0.5cm}
\end{figure}
Fig.~\ref{fig:cap_doppler_quadriga}~and~\ref{fig:tp_doppler_quadriga} show the sum-rate and throughput performances of RSMA and SDMA with respect to user speed. We analyze the scenarios with $n_{t}=32$, $K=8$ and average received SNR levels of $25$dB and $35$dB in the simulations. 
We start by comparing the sum-rate performance achieved by SDMA and RSMA using the power allocation coefficient obtained from the expression \eqref{eq:poweralloc} and the coefficient that maximizes the lower bound \eqref{eqn:lowerbound}, which is found by exhaustive search over the power allocation coefficient $t$. For the exhaustive search procedure, we span the interval of $t \in \left(0,1 \right]$ with a granularity of $0.001$. It can be observed from Fig.~\ref{fig:cap_doppler_quadriga} that the power allocation coefficients obtained from the closed-form solution \eqref{eq:poweralloc} can achieve a performance close to the coefficients obtained by the exhaustive search for all user speeds and both SNR values.  
One can observe from Fig.~\ref{fig:cap_doppler_quadriga}~and~\ref{fig:tp_doppler_quadriga} that RSMA achieves a robust performance and reliable multi-user communications with the proposed power allocation algorithm under user mobility. The sum-rate and achieved throughput of SDMA drop significantly with increasing user speed and saturate with increasing SNR. On the other hand, RSMA operates with less performance degradation and achieves a non-saturating robust performance as the user speed increases. 

Next, we extend our analysis to a system model with multi-antenna receivers, {\sl i.e.,} $n_{r}$ antennas per receiver. A vector of $Q_{c}$ common streams and $K$ vectors of $Q_{p}$ private streams are transmitted, resulting in $Q_{c}+ KQ_{p}$ total transmitted streams. The precoders for the private streams are obtained by the Block-Diagonalization (BD) method \cite{spencer_2004}.  
We employ the power allocation algorithm in \eqref{eq:poweralloc} without any modifications. The power allocated to the common precoders is distributed equally among the common streams and the power allocated to the private precoders is distributed equally among the private streams of all users.
Fig.~\ref{fig:tp_doppler_qc2qp2_quadriga}~and~\ref{fig:tp_doppler_qc3qp3_quadriga} show the throughput results for $n_{t}=64$, $n_{r}=4$, $K=8$ and different values of $Q_{c}$ and $Q_{p}$. The throughput of RSMA is observed to have a non-monotonic behaviour, as the proposed power allocation algorithm is not optimal for multi-antenna receivers. The figures show that RSMA with the proposed power allocation algorithm performs significantly better than SDMA with multi-antenna receivers as well. We refer the interested reader to \cite{dizdar_2021} for a more involved numerical analysis.

The numerical results show that RSMA is robust to the effects of user mobility and can achieve significantly higher sum-rate and throughput performance than SDMA, for which the performance saturates with increasing SNR. We conclude that RSMA can solve the mobility problem and meet the demand for higher data rate services in 6G. 

\section{Conclusion}
In this paper, we discussed RSMA in the light of requirements and emerging problems for enhanced URLLC and eMBB in 6G. We present the optimal system designs employing RSMA that target low-latency communications as well as high-throughput and robust communications under user mobility and latency in the network. We demonstrate that RSMA achieves robust and reliable performance under the considered target scenarios, and thus, is an excellent enabling technology for 6G. Future work includes the performance demonstration of RSMA by real transceiver prototypes.

\vspace{12pt}
\color{red}

\end{document}